\documentclass[useAMS]{mn2e}
\usepackage{amssymb,amsmath,psfig,times}
\usepackage{graphicx}
\def\gsim{ \lower .75ex \hbox{$\sim$} \llap{\raise .27ex \hbox{$>$}} }
\def\lsim{ \lower .75ex\hbox{$\sim$} \llap{\raise .27ex \hbox{$<$}} }

\def\beq{\begin{equation}}
\def\eeq{\end{equation}}

\def\fe{{\it Fermi}}
\def\ba{BATSE}

\def\epo{$E^{\rm obs}_{\rm p}$}

\title[Photospheric emission in GRB 100507]
{Photospheric emission throughout GRB 100507 detected by {\it Fermi}}
\author[G. Ghirlanda et al.]
{G. Ghirlanda$^{1}$\thanks{E--mail:giancarlo.ghirlanda@brera.inaf.it}, 
A. Pescalli$^{2}$, G. Ghisellini$^{1}$\\
$^{1}$INAF -- Osservatorio Astronomico di Brera, via E. Bianchi 46, I-23807 Merate, Italy\\
$^{2}$Dipartimento di Fisica G. Occhialini, Universit\'a di Milano Bicocca, Piazza della Scienza 3, I-20126 Milano, Italy\\
}
\begin{document}

\date{Accepted 2013 April 17.  Received 2013 April 2; in original form 2013 February 18}


\maketitle

\label{firstpage}

\begin{abstract}
Gamma ray bursts with blackbody spectra are only a few  and in most cases this spectral 
component is accompanied by a dominating non--thermal one. Only four bursts detected by \ba\ have a pure
blackbody spectrum throughout their duration. We present the new case of GRB 100507 detected by the Gamma Burst Monitor on board the \fe\ satellite. GRB 100507 has a  blackbody spectrum for the entire duration ($\sim$30 s) of the prompt emission. The blackbody temperature varies between  25 and 40 keV. The flux varies between $10^{-7}$ and $4\times10^{-7}$ erg/cm$^2$ s. There is no clear evidence of a correlation 
between the temperature and the blackbody flux. If the thermal emission in GRB 100507 is due to the fireballs becoming transparent, we can estimate the radius $R_T$ and bulk Lorentz factor $\Gamma_T$ corresponding to this transition and the radius $R_0$ where  the fireballs are created. We compare these parameters with those derived for the other four bursts with a pure blackbody spectrum. In all but one burst, for fiducial assumptions on  the  radiative efficiency and distance of the sources, $R_0\sim10^{9}-10^{10}$ cm, i.e. much larger than the  gravitational radius of a few solar mass black hole. Possible solutions of this apparent inconsistency are tentatively discussed considering the dependence of $R_0$ on the unknown parameters.  Alternatively, such a large $R_0$ could be where the fireball, still opaque, converts most of its kinetic energy into internal energy (due to the impact with some material left over by the progenitor star) and starts to re--accelerate.   
\end{abstract}
\begin{keywords}
Gamma-ray: bursts, radiation mechanisms: thermal
\end{keywords}

\section{Introduction}

Although the prompt emission spectrum of Gamma Ray Bursts (GRBs) is typically 
non--thermal, as expected if their emission is produced by shock accelerated relativistic electrons 
which radiate via synchrotron/inverse Compton, there are a few bursts which show a thermal 
blackbody (BB) spectrum. Thermal emission\footnote{Thermal emission is used for ``blackbody emission" throughout the paper.} is expected in the so called ``standard" fireball model 
of GRBs when the relativistically expanding plasma  becomes transparent 
(e.g. Goodman 1986; Paczynski 1986; Daigne \& Mochkovitch 2002). Also alternative GRB 
models like the ``fireshell" model (Ruffini et al. 2004; Bernardini et al. 2005)  or the ``cannonball" model 
(Dar \& de Rujula 2004) predict a thermal spectral component in GRBs. Thermal photons can be either those 
of the initial fireball or they can be created at some stage of the fireball evolution, when it is still moderately 
opaque, due to some dissipation mechanism (e.g. Rees \& Meszaros 2005). 

Time resolved spectroscopy of bright bursts, 
detected by the Burst And Transient Source Experiment (\ba) on board the Compton Gamma Ray Observatory,   
revealed a thermal component either at the 
beginning of GRB 910807, 910927, 911118, 970111, 980306 (Ghirlanda, Celotti \& Ghisellini 2003 -- G03) or throughout 
the entire burst duration in GRB 930214, 941023, 951228 (Ryde 2004 -- R04) and GRB 990413 
(Bosnjak, Celotti \& Ghirlanda 2006 -- B06). To date these are the only bursts in which a ``pure" blackbody 
component is observed. It has been shown (Ryde et al. 2005 - 
but see Ghirlanda et al. 2007) that in a larger fraction of \ba\ bursts, time resolved spectra
could be modelled  as the superposition of a  BB (contributing a minor fraction of the total flux) 
and a dominating non--thermal component which is typically a single power law in the \ba\ energy range 
($\sim$30 keV -- 1 MeV). 

\begin{figure}
\psfig{file=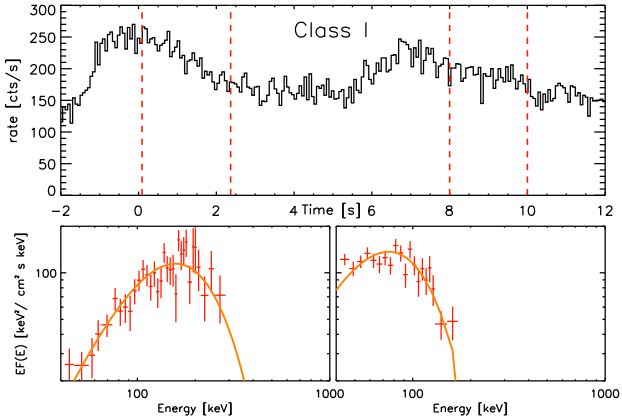,width=8.5cm}
\vskip -0.3 cm
\caption{Example of {\it Class--I} bursts. GRB 951228 (R04) which has a thermal spectrum throughout its duration. Top panel: light curve (not background subtracted). Bottom panels: two spectra (accumulated in the time intervals marked by the dashed vertical lines in the top panel). Both spectra are fitted with a Black Body component (solid orange lines in the two bottom panels).  }
\label{fg00}
\end{figure}
\begin{figure}
\psfig{file=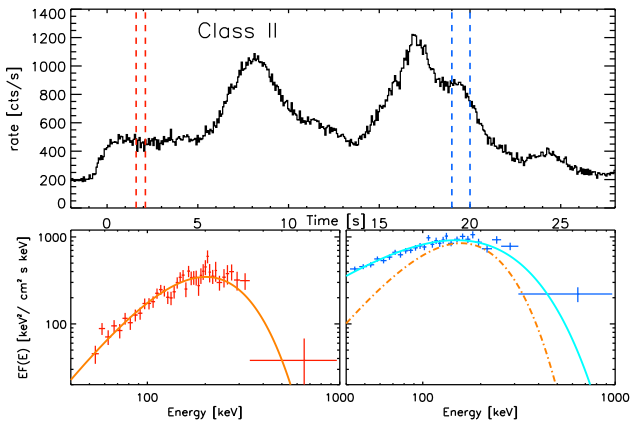,width=8.5cm}
\vskip -0.3 cm
\caption{Example of {\it Class--II} bursts. GRB 970111 (G03)  has a thermal spectrum at the beginning and a non--thermal spectrum at later times. Top panel: light curve (not background subtracted). Bottom panels: two spectra (accumulated in the time intervals marked by the dashed vertical lines in the top panel). The first spectrum is fitted with a blackbody  (solid orange line) the second spectrum is fitted by a non--thermal component (a cutoff--powerlaw model in this case) shown by the solid cyan line. For comparison the BB spectrum fitted to the early time spectrum (normalized to the peak of the late time spectrum) is shown by the dash--dotted orange line.}
\label{fg01}
\end{figure}

\begin{figure}
\psfig{file=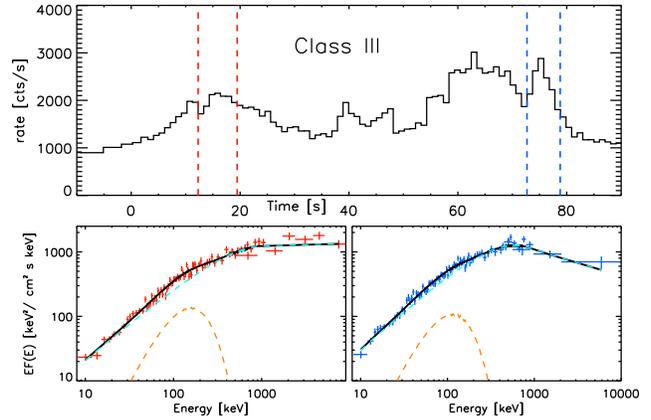,width=8.5cm}
\vskip -0.3 cm
\caption{Example of {\it Class--III} bursts. GRB 100724 (Guiriec et al. 2012) which has a thermal component (contributing only the 5\% of the total flux) and a dominating  non--thermal component throughout its duration. Top panel: light curve. Bottom panels:  early and late time spectrum (corresponding to the time intervals of the dashed vertical lines in the top panel) deconvolved into a non--dominating BB component (dashed orange line) and a Band function (dashed cyan line). The total spectrum is shown by the solid black line.}
\label{fg02}
\end{figure}

Recently, GRB 110721A (Axelsson et al. 2012), GRB 100724B (Guiriec et al. 2011) and GRB 120323A (Guiriec et al. 2012), detected by the \fe\ satellite, showed the compresence of a BB, contributing $\sim$5\% of the total flux,  and a dominating non--thermal component. GRB 090902B (Ryde et al. 2010), also detected by \fe,  has a prominent ``broadened" BB component and a non--thermal sub--dominant power law. 

Therefore, evidences of thermal components in GRB prompt spectra could be divided into three classes as follows. 
\begin{enumerate}
\item {\it Class--I} bursts with a pure BB component for the entire duration of their prompt emission phase. These are only four cases and they have all been detected by \ba\ (R04, B06). An example of this class is shown in Fig.\ref{fg00}.
\item {\it Class--II} bursts with a pure BB component only in the first few seconds of their emission which is overtaken by a dominating non--thermal component afterwards (G03). An example is shown in Fig.\ref{fg01}.
\item {\it Class--III} bursts with a BB plus a non--thermal component (either a power law or a more complex double power law - e.g. Ryde 2005, Axelsson 2012, Guiriec 2011, 2012) throughout their duration. An example of this class is shown in Fig.\ref{fg02}.
\end{enumerate}

We do not pretend to define a scheme for classifying GRB spectra, which show complex evolutions. GRB 090902B, for example, could be of class--I for its remarkably dominating broadened BB component, but the presence of an underlying non thermal component is also typical of class--II events. The above classification scheme is used as a guideline for helping the discussion of the results presented in this paper.


A common feature\footnote{In GRB 100724B (Guiriec et al. 2012) shown in Fig.\ref{fg02}, belonging to class--III, the temperature of the thermal component is almost constant throughout the burst.} of the thermal component is that its temperature evolves with time approximately as $kT\propto t^{-2/3}$ (R04) or $kT\propto t^{-1/4}$ (G03) after an initial rising or constant phase. Also the flux of the thermal component decreases $\propto t^{-2}$ at late times. Such temporal behaviours at late times have been interpreted (Peer 2008) as high latitude emission from the optically thick surface of the expanding plasma when it becomes transparent. 

Possible interpretations of class--II and III bursts propose that the non--thermal component is produced by Compton scattering of the photospheric photons by relativistically accelerated electrons (e.g. Peer 2008) or that the thermal component is the photospheric emission and the non--thermal component is produced in the optically thin region (e.g. through internal shocks) so that the relative strength of these two components is regulated by the thermal/magnetic content of the jet (Hascoet, Diagne \& Mochkovitch 2013). 


The  few GRBs with only a ``pure" blackbody spectrum represent a fundamental tool to investigate the origin of the thermal component, its evolution and the basic properties of the relativistic outflow. The detection of a blackbody spectrum in GRBs, if produced by the relativistic outflow becoming transparent, allows us to estimate (Peer et al. 2007) the transparency radius $R_{\rm T}$, the bulk Lorentz factor $\Gamma_{\rm T}$ at transparency and the radius at the base of the relativistic flow $R_0$  where the acceleration began. The latter,  in the standard picture, should be a few times the gravitational radius $r_{\rm g}$ of a few solar mass black hole (BH; i.e. the putative central engine of GRBs). Clearly, when only the thermal component is observed (class--I) these estimates depend on a lower number of free parameters (since no modelling of the non--thermal component is required in these cases). Interestingly,  it should be explained why in class--I bursts the non--thermal emission (typically ascribed to synchrotron/inverse Compton emission at internal shocks) is absent: either it could be highly inefficient or there should be a mechanism suppressing this process. 


In this paper we present the spectral analysis of the prompt emission of GRB 100507 as the first GRB detected by \fe\ which shows a pure BB spectrum throughout its duration (\S3). This burst increases the number of known class--I events to five.  The spectral evolution of this burst (\S4) is compared to that of the other four class--I bursts. Within the standard fireball model where the observed thermal component is interpreted as photospheric emission (\S5),  we estimate $R_{\rm T}$, $\Gamma_{\rm T}$ and $R_0$ (\S6)  and discuss our results in \S7. Throughout the paper we assume a standard flat Universe  with $h=\Omega_{\Lambda}=0.7$.

\section{Selection of candidates ``thermal" bursts}

Most GRB {\it time integrated spectra} are fitted with the Band function (i.e. two power laws joined by a smooth rollover - Band et al. 1993) or by a simpler power law with a high energy exponential cutoff (CPL hereafter). Both these models have a low energy power law component $N(E)\propto E^{\alpha}$ [where $N(E)$ represents the photon spectrum]. For particularly hard low energy spectra, i.e. when $\alpha\simeq 1$, the CPL function approximates  the Rayleigh--Jeans portion of a Planck function (i.e. $EN(E)\propto E^2$ for $E\ll kT$). Therefore,  bursts with a hard low energy time integrated spectrum are good candidates to have a thermal component. 

The five \ba\ GRBs studied in G03 had a time integrated spectrum with a photon spectral index $\alpha\ge0$. The {\it time resolved}  analysis of these bursts revealed that their spectra, during the first few seconds of the emission, were consistent with a blackbody while later  the spectrum became substantially non--thermal (i.e. a CPL with $\alpha<0$). One of these bursts, representative of class--II events, is shown in Fig.\ref{fg01}. The spectral evolution, from an initial thermal spectrum (lasting only a small fraction of the burst duration) to a longer--lived non--thermal spectrum, explains why the time integrated spectrum of these bursts have $\alpha>0$.  It is also expected that if some dissipation mechanism is operating when the fireball is still opaque (i.e. below the photosphere) the thermal spectrum released at the photosphere should have a low energy slope softer than the photon limit $\alpha=1$ of the blackbody (e.g. Peer 2008, Beloborodov 2010). In class--III bursts (Fig.\ref{fg01}) the time integrated spectrum is dominated by a non--thermal component and only a time resolved analysis can reveal the non--dominating BB component of their spectra. 

Therefore, a successful way to identify  thermal components  in GRBs would require a systematic time resolved spectral analysis. In this paper  we are interested in class--I  bursts, i.e. those with a blackbody spectrum throughout their duration, which should have a time integrated spectrum with $\alpha>0$. 


We consider the spectral catalog of GRBs detected by the Gamma Burst Monitor (GBM) on board the \fe\ satellite. The GBM spectral catalogue (Goldstein et al. 2012, see also Nava et al. 2011) contains 487 GRBs detected during the first two years of the mission (2008 July 14 -- 2010 July 13) with enough flux to perform a spectral analysis. Both the time integrated spectra and the spectra accumulated at the peak of the light curve of individual bursts are analysed with different spectral functions (Goldstein et al. 2012). We  selected the bursts with a hard low energy spectrum, i.e. $\alpha>0$. We find 10 bursts satisfying this selection: four have a time integrated spectrum with $\alpha>0$ and the remaining six have a peak spectrum with $\alpha>0$. However, in all but one of these bursts the values of $\alpha$ reported in the catalogue have large uncertainties which make them consistent with $\alpha\le0$. 

In this paper, we present  GRB 100507 which has a  low energy (time integrated) spectral index ($\alpha=0.41\pm0.09$)  $\sim 4.4\sigma$ harder than the $\alpha=0$ limit\footnote{For completeness, we also analszed the other selected bursts, but the extracted spectra had too few counts to perform a reliable time resolved spectral analysis.}. GRB 100507 is a long burst ($T_{90}\simeq 35$ s) with a variable light curve with multiple spikes (top panel of Fig.\ref{fg1}). It has a time integrated spectrum fitted by a CPL model with a peak energy of the $\nu F(\nu)$ spectrum \epo=137.4$\pm$4.1 keV and a fluence (integrated in the 8--1000 keV energy range) of 7.2$\times 10^{-6}$ erg s$^{-1}$ (Goldstein et al. 2012).  

\section{Time resolved spectral analysis}
The GBM (Meegan et al. 2009) comprises 12 thallium sodium iodide [NaI(Tl)] and two bismuth 
germanate (BGO) scintillation detectors which cover the energy ranges  $\sim$8 keV--1 MeV 
and $\sim$300 keV--40 MeV, respectively. The GBM acquires different data types for the spectral analysis 
(Meegan et al. 2009). Data files and detector response files for individual bursts are publicly available 
in a dedicated archive\footnote{http://heasarc.gsfc.nasa.gov/W3Browse/fermi/fermigbrst.html}.
\begin{table}
\begin{center}
\begin{tabular}{ccccccccc}
\hline\hline
  \multicolumn{1}{c}{$t_{\rm start}$} &
  \multicolumn{1}{c}{$t_{\rm stop}$} &
  \multicolumn{1}{c}{$kT$ (keV)} &
  \multicolumn{1}{c}{$F$ (erg/cm$^{2}$s)} &
  \multicolumn{1}{c}{C-stat/DOF} \\
\hline
  -3.50 & 1.75 & 38.02$\pm$2.37 & (1.44$\pm$0.11)E-7 & 1.20\\
  1.75 & 2.62 & 31.97$\pm$2.09 &  (2.62$\pm$0.23)E-7 & 1.12\\
  2.62 & 4.37 & 34.43$\pm$2.10 & (2.23$\pm$0.17)E-7 & 1.11\\
  4.37 & 5.25 & 39.25$\pm$2.47 & (3.72$\pm$0.30)E-7 & 1.06\\
  5.25 & 7.00 & 40.62$\pm$2.44 & (2.58$\pm$0.20)E-7 & 1.00\\
  7.00 & 10.50 & 30.63$\pm$1.34 & (1.72$\pm$0.10)E-7 & 1.05\\
  10.50 & 12.25 & 40.43$\pm$2.39 & (2.80$\pm$0.21)E-7 & 1.20\\
  12.25 & 14.00 & 34.53$\pm$1.48 & (3.58$\pm$0.20)E-7 & 1.07\\
  14.00 & 17.50 & 31.52$\pm$1.05 & (2.94$\pm$0.13)E-7 & 1.28\\
  17.50 & 21.00 & 26.26$\pm$0.88 & (2.28$\pm$0.10)E-7 & 1.09\\ 
  21.00 & 28.00 & 28.72$\pm$1.13 & (1.33$\pm$0.07)E-7 & 1.31\\
\hline\hline
\end{tabular}
\caption{Time resolved spectral evolution of GRB 100507. $t_{\rm start}$ and $t_{\rm stop}$  are referenced to the trigger time of the burst and represent the time interval where each time resolved spectrum is accumulated. The blackbody temperature, flux (integrated in the 10 keV -- 1MeV energy range) and the Cash statistic/degrees of freedom are reported. Errors are at the 90\% confidence level. }
\label{tab1}
\end{center}
\end{table}

For the time resolved spectral analysis we used the public software 
RMFIT\footnote{http://fermi.gsfc.nasa.gov/ssc/data/analysis/user/} (\texttt{v3.3pr7}). 
In order to model the background spectrum for the time resolved spectral analysis, we 
selected two time intervals before and after the burst. The sequence of background spectra in the 
two selected intervals were fitted with a first order polynomial to account for 
the possible time variation of the background spectrum. Then the background spectrum was 
extrapolated to the time intervals selected for the time resolved spectroscopy. 

We consider the two NaI (\#10, 11) detectors (8 keV -- 1MeV) that were triggered by the burst\footnote{We verified that adding also other NaI detectors, which had a lower signal relative to the two most illuminated ones considered does not change our results}. The signal 
in the two BGO detectors (300 keV -- 40 MeV) is too low to perform a time resolved spectral analysis. A free 
normalization constant was considered in fitting simultaneously the two detectors' spectra with the same 
spectral model. 

\begin{figure}
\psfig{file=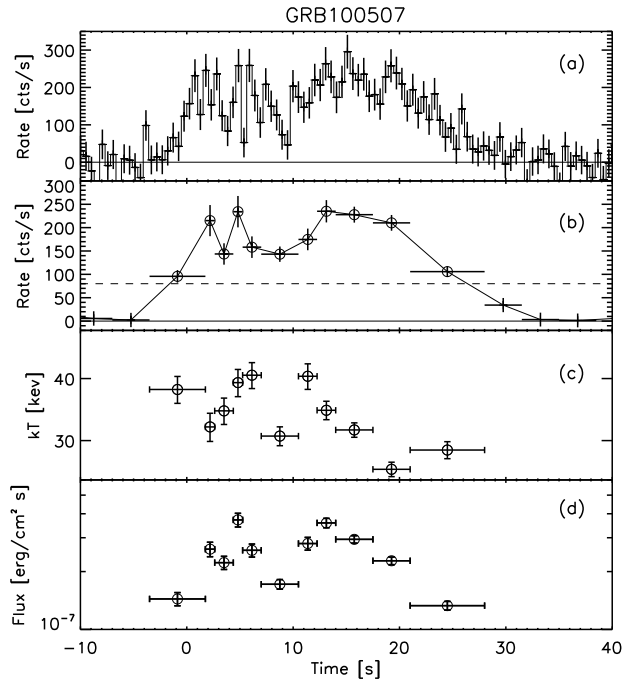,width=8.5cm}
\caption{Spectral evolution of GRB 100507 fitted with a blackbody model. Panel (a) shows the  count rate light curve (background subtracted) with a time resolution of 0.512 s from detector NaI \#10. Panel (b) shows the light curve from the same detector rebinned with a variable time resolution. The analysed spectra are those exceeding the 80 counts/s level (dashed line). Panels (c) and (d) show the time evolution of the BB temperature $kT$ and of the BB flux (integrated in the 1--1000 keV energy range).}
\label{fg1}
\end{figure}

\begin{figure}
\hskip -1 cm
\includegraphics[width=0.6\textwidth]{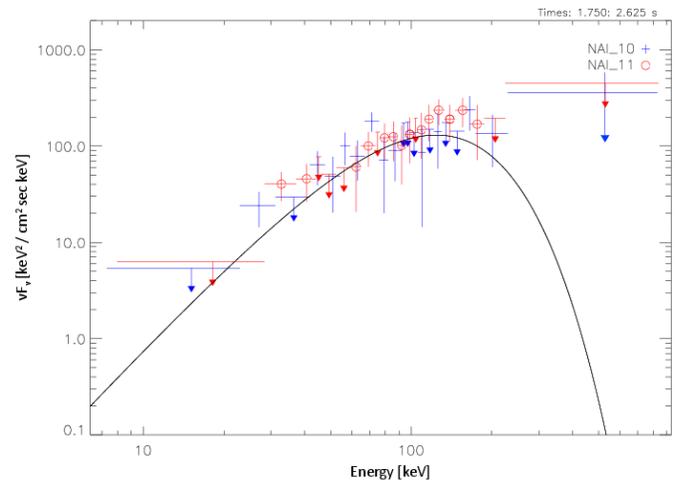}
\caption{Example of a spectrum (corresponding to the time interval 1.75--2.625s) of GRB 100507 fitted with a BB model (solid line). The two data sets are of the two NaI considered in the analysis (upper limits are at 3$\sigma$).  }
\label{fg11}
\end{figure}

Fig.\ref{fg1}a shows the count rate light curve (of detector NaI\#10) with a 0.512 s resolution. We rebinned 
the light curve in time extracting 11 time resolved spectra (Fig.\ref{fg1}b), 
so as have a minimum rate of 80 counts s$^{-1}$ in each time resolved spectrum. This request ensures to 
have enough counts in the single spectra to constrain the spectral parameters of the fitted model.  
The extracted spectra have a signal to noise ratio between 11 and 22. 
We show as an example the second time resolved spectrum fitted by a BB model in Fig.\ref{fg11}. 

First we fitted the 11 time resolved spectra with the typical Band function (Band et al. 1993) used for time--integrated spectra of GRBs. It has been recently suggested that the Band function could have a thermal origin in collisional heated jets (Beloborodov 2010) and that typical low and high energy spectral slopes (0.4 and -2.5 respectively) should be expected. In all time resolved spectra we could not constrain the high energy power law spectral index of the Band function. Moreover, the resulting low energy photon spectral index was found to be hard ($>$0) but with an associated large uncertainty. This suggests that these spectra could be better fitted with a blackbody model which has also less free parameters.

Therefore, 
we fitted the 11 spectra with a blackbody (BB). The evolution of the temperature $kT$ and of the flux (integrated in the 10--1000 keV energy range) is shown in Figs. \ref{fg1}(c) and (d), respectively. The fit with the BB model is adequate for all the considered spectra; the spectral parameters  are reported in Tab.\ref{tab1}. 
GRB 100507 has a spectrum well described by a blackbody for all the duration of the burst.

\section{Spectral evolution}

\begin{figure}
\hskip -1 cm
\includegraphics[width=0.55\textwidth]{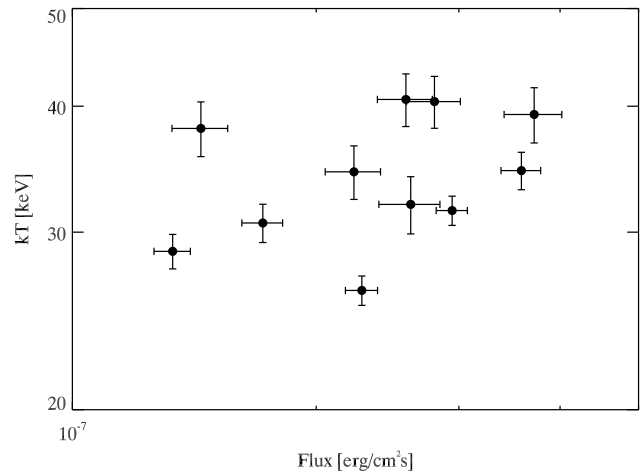}
\caption{Temperature versus flux of the blackbody fit to the time resolved spectra (Table \ref{tab1}). }
\label{fg1a}
\end{figure}

The \ba\ bursts with a BB component show a monotonic evolution of the temperature $kT$ with time: a possible initial phase of constant temperature (G03) or a mildly increasing temperature (R04) is typically followed by a  decrease  with time $kT\propto t^{-1/4;-2/3}$. In these bursts the flux seems to behave  similarly, being strongly correlated with the temperature $kT$. 
 
In GRB 100507 there is no evidence of a strong correlation between the temperature and the flux, as shown in Fig.\ref{fg1a}. The Spearman correlation coefficient is 0.4 with a high chance correlation probability, i.e. 0.2.

The temperature evolution in GRB 100507  does not change monotonically during the burst and remains between $\sim$30 and 40 keV (Fig.\ref{fg1}c). Also in the class--III GRB 100724B (Guiriec et al. 2012) the temperature of the thermal component does not vary monotonically as in other thermal bursts and changes within a narrow range. Also in the \fe\ GRB 090902B (Ryde et al. 2010) $kT$ of the multi--temperature blackbody component does not follow the typical time decay observed in several \ba\ bursts. 

The values of $kT$ derived for GRB 100507 are higher than the low energy threshold (8 keV) of the GBM instrument, ensuring that in these spectra  we are measuring the full curvature of the Plank spectrum 
\footnote{This was not the case of two of the three \ba\ bursts (GRB 941023, 951228) claimed to have a thermal spectrum throughout their duration (R04)}. GRB 100507 is the first burst of class--I detected by \fe. Compared to other class--I events, GRB 100507 is peculiar also because its the BB flux  is poorly correlated with the temperature and varies only by  a factor of $\sim$4 during the burst (Fig.\ref{fg1}b). 

\section{Photospheric emission}

In the standard fireball scenario the emission of thermal radiation is expected when the fireball becomes transparent. 
After an initial acceleration phase due to the internal pressure of the photons (and/or to the magnetic field) the outflow optical depth becomes $\tau\leq 1$ and the internal radiation is emitted. This emission should have a blackbody spectrum, unless some dissipation process, below the photosphere, modifies the spectral energy distribution of the trapped--in photons (e.g. Peer 2008, Beloborodov 2010). 

The comoving frame luminosity (isotropic equivalent) of the photons trapped within the fireball is $L'=4\pi R^{2} \sigma T'^{4}$, where $T'$ is the comoving temperature of these photons. Adopting the transformation from the comoving to the rest frame, i.e. $L=4/3 \Gamma^{2} L'$ and $T=5/3 \Gamma T'$ (Ghirlanda et al. 2012), we can derive the ratio between the fireball radius and its bulk Lorentz factor at the transparency:
\begin{equation}
\frac{R_T}{\Gamma_T}=2.406 \frac{d_{L}(z)}{(1+z)^2}\left(\frac{F_{B}}{\sigma T_{B,o}^4}\right)^{1/2} \,\,\,\, {\rm cm}
\label{eq1}
\end{equation}
where $F_{B}$ and $T_{B,o}$ are the flux and the observer--frame temperature of the blackbody spectrum and $z$ is the source redshift. If we observe a black--body spectrum, we can relate the observables (i.e. $F_{B}$ and $T_{B,o}$) to the ratio $R_T/\Gamma_T$.

The transparency  can be reached:
\begin{enumerate}
\item during the {\it acceleration} phase, i.e. when the fireball is still accelerating with a bulk Lorentz factor $\Gamma\propto R$. Equation \ref{eq1} allows us to derive $R_{0}/\Gamma_{0}$ and, assuming that the fireball is created at $R_{0}$ with $\Gamma_{0}=1$, this gives the radius at the base of the jet where the fireball is created. In this case we should observe the temperature of the fossil photons, i.e. those that were trapped in at $R_{0}$;

\item during the {\it coasting} phase, i.e. when most of the internal energy has been converted to bulk motion, i.e. in the so called ``coasting" phase when $\Gamma=$const.  In this case, the transparency radius is (Daigne \& Mochkovitch 2002):
\begin{equation}
R_{T}=\frac{L_{0}\sigma_{T}}{8\pi m_{p}c^{3}\Gamma^{3}} \,\,\,\,\,\,\, \, \, {\rm cm}
\label{eq2}
\end{equation} 
where $\sigma_{T}$ is the Thomson cross section and $c$ the speed of light. $L_{0}$ represents the initial total luminosity injected by the central engine in the fireball at $R_{0}$. Only a fraction  $L\sim \eta L_{0}$ with $\eta<1$ is released as radiation at the transparency. A typical value $\eta\sim20\%$ is derived from the modeling of GRB afterglow lightcurves.  From equations \ref{eq1} and Eq.\ref{eq2}, the transparency radius $R_{T}$ and corresponding bulk Lorentz factor  $\Gamma_{T}$ result:
\begin{equation}
R_{T}=1.624 \left(\frac{\sigma_{T}}{m_{p}c^{3}\eta}\right)^{1/4} \frac{d_{L}(z)^{5/4}}{(1+z)^{3/2}}\frac{F_{B}^{5/8}}{(\sigma T_{B,o}^4)^{3/8}} \,\,\,\,\,\,\, {\rm cm}
\label{eq3}
\end{equation}
\begin{equation}
\Gamma_{T}=0.675 \left(\frac{\sigma_{T}}{m_{p}c^{3}\eta}\right)^{1/4} [d_{L}(z)(1+z)^{2}]^{1/4} F_{B}^{1/8}(\sigma T_{B,o}^4)^{1/8}
\label{eq4}
\end{equation}
Finally, the energy density of the fossil BB photons (injected at $R_{0}$) is $aT_{0}^{\prime 4}=L_{0}/4\pi c R_{0}^2$, where $a$ is the radiation constant. Through the scaling relations valid for the acceleration phase (i.e. $\Gamma\propto R$ and $T\propto R^{-1}$) and for the coasting phase (i.e. $\Gamma$=const and $T\propto R^{-2/3}$), it is possible to derive the radius $R_{0}$ where the fireball is created if the transparency is reached during the coasting phase: 
\begin{equation}
R_{0}=12.5\, \eta^{3/2} \frac{d_{L}(z)}{(1+z)^{2}} \left(\frac{F_{B}}{\sigma T_{B,o}^4}\right)^{1/2} \,\,\,\,\,\,\,\, {\rm cm}
\label{eq5}
\end{equation}
 
\end{enumerate}


\begin{figure*}
\hskip 1cm
\psfig{file=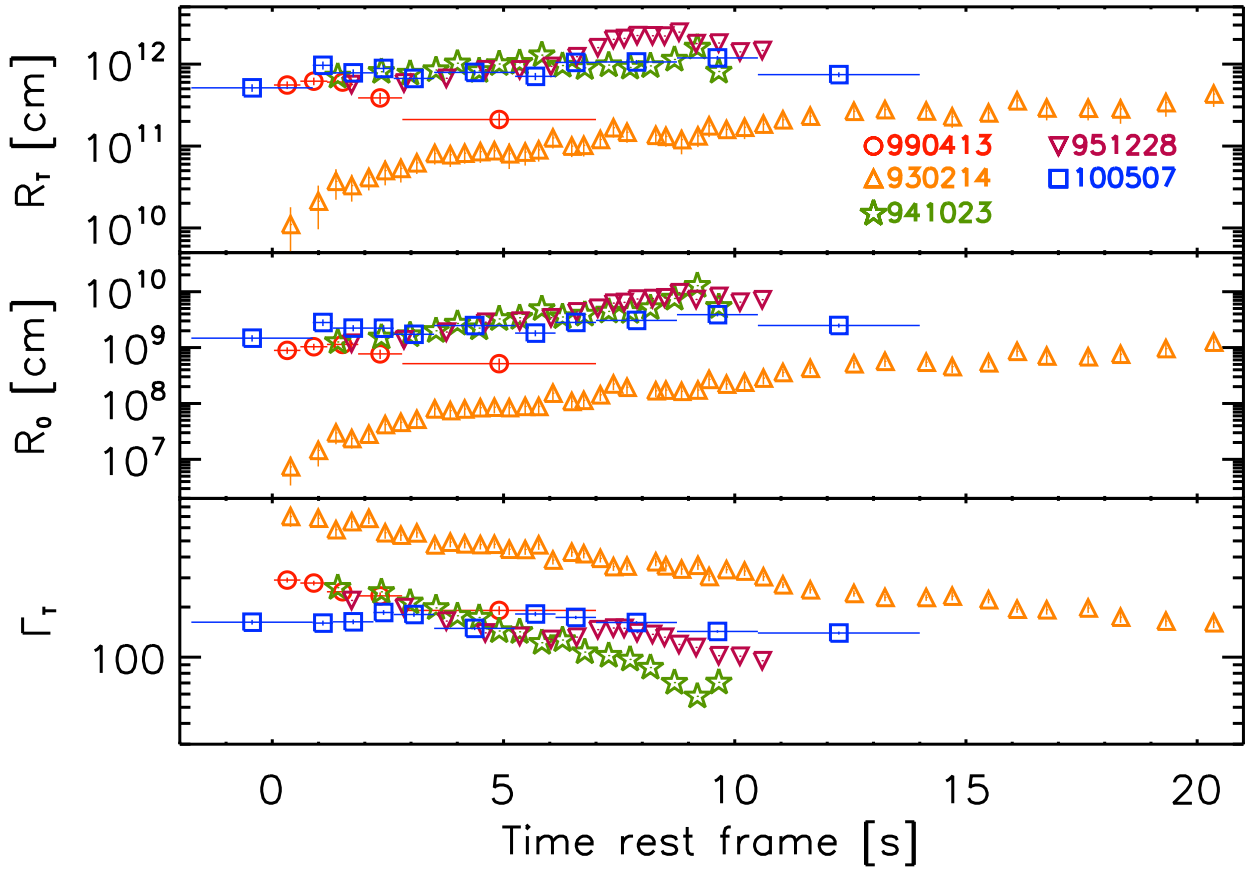,width=17cm}
\caption{Class--I bursts fireball parameters evolution. From top to bottom: evolution of the transparency radius $R_T$,  of the radius of formation of the fireball $R_0$ and of the bulk Lorentz factor $\Gamma_T$ at the transparency radius. $z=1$ and $\eta=20$\% have been assumed for all bursts. The different symbols show the five GRBs with only a thermal dominating blackbody component in their time resolved spectra from the beginning to the end. The triangles, stars and upside--triangles show GRB 930214, 941023 and 951228, respectively (Ryde 2004; Ryde \& Peer 2005), the circles are for GRB 990413 (B06) and the squares for GRB 100507 (this paper). }
\label{fg2}
\end{figure*}


\section{Estimate of fireball parameters} 

If a blackbody spectrum is observed,  the measurement of its temperature $kT_{B,o}$ and flux $F_{B}$ allow us  to estimate some fundamental parameters of the fireball. We consider the five class--I GRBs with a clear evidence of a blackbody spectrum throughout their duration. In addition to the \ba\ GRBs 930214, 941023, 951228 (R04) and 990413 (B06), whose spectral parameters have been collected from the literature, we  add in this paper the \fe\ GRB 100507. 

The equations of $\Gamma_{T}$, $R_{T}$ and $R_{0}$ reported in \S5 depend on the distance of the source through the luminosity distance $d_{L}(z)$ and the term $1+z$. Since $z$ is unknown for the five class--I bursts we assume $z=1$, i.e. typical of long duration GRBs. The dependence of our estimates on $z$ are discussed in \S7.

First we consider case (ii), i.e. if the fireball becomes transparent during the coasting phase,  through equations \ref{eq3}--\ref{eq5}, we can derive $R_{T}$, $R_{0}$ and $\Gamma_{T}$ which are shown in Fig.\ref{fg2} (from top to bottom respectively). In these estimates we have assumed that the radiative efficiency at the photosphere is $\eta$=20\%. We note that four of the five bursts considered have similar values of the three parameters and similar temporal evolution. In GRB 941023, 951228, 990413 and 100507 the  values of the transparency radius $R_{T}\sim 10^{12}$ cm (top panel in Fig.\ref{fg2}) are consistent with what expected for the typical photospheric radius of GRBs (e.g. Daigne \& Mochkovitch 2002). The bulk Lorentz factors $\Gamma_T$ (bottom panel in Fig.\ref{fg2}) for these four bursts are between 50 and 200 and, in at least two cases, $\Gamma_{T}$ seems to decrease monotonically with time. In GRB 990413 and 100507, instead, $\Gamma_T$ is almost constant $\simeq$200. Also these values of $\Gamma_{T}$ are consistent with the recent estimates of this parameter obtained through the modelling of the optical afterglow light curves of a sample of 30 GRBs (Ghirlanda et al. 2012). The initial fireball radius $R_0$ (middle panel of Fig.\ref{fg2}) is in the range $10^{9}-10^{10}$ cm for the four GRBs with similar evolution. We assumed that the fireballs have $\Gamma_0=1$ at $R_0$. 

GRB 930214 shows a different behaviour: the transparency radius evolves from $R_{T}\sim10^{10}$ cm at the beginning to 5$\times 10^{11}$ cm at the end of the burst.  The evolution of $\Gamma_T$ spans almost one order of magnitude, starting  with $\Gamma_{T}\sim 700$ and reaching $\sim$100 at the end of the burst. The radius $R_0$ is initially smaller than what estimated for the other four bursts and it varies by two orders of magnitudes during the $\sim$20 s of duration of the burst. However, also in GRB 930214, $\Gamma_T$ shows a very small amplitude variability (i.e. much smaller than a factor of 2).

If the transparency is reached during the acceleration phase (when  $\Gamma\propto R$) through equation\ref{eq1} and assuming  $\Gamma_0=1$ it is possible to estimate $R_{0}$ which would have the same dependence from the variables as in equation \ref{eq5} except that it is independent from $\eta$. The values of $R_{0}$ shown in Fig.\ref{fg2} (mid panel) that were derived through equation \ref{eq5} under the assumption that the transparency is reached during the coasting phase, would increase by a factor of $\sim0.46\eta_{0.2}^{3/2}$ if the transparency is reached during the acceleration phase. 

\section{Discussion}

Either if the transparency is reached during the coasting phase or the acceleration phase, we find that the radius $R_{0}$ where the fireball is created is of the order of $10^9-10^{10}$ cm (Fig.\ref{fg2} mid panel) except for the GRB 930214 where $R_{0}$ increases  two orders of magnitudes during the burst.
 
\subsection{Standard fireball model}
In the standard fireball scenario, the values of $R_{0}$ derived in \S6 should be comparable with the typical  gravitational radius $r_g$ of a few solar masses black hole. Although we do not know exactly the details of the jet formation, we should expect that $R_{0}\simeq10r_{g}$.  For a black hole of mass $M_{bh}=5M_{\odot}$ the values of $R_{0}$ shown in Fig.\ref{fg2} for the bursts of class--I are a factor of $10^2-10^3$ larger than  $10 r_{g}$. In the following we discuss how the  estimated values of $R_{0}$ can be reduced by changing the assumed values of some unknown parameters:
\begin{figure}
\psfig{file=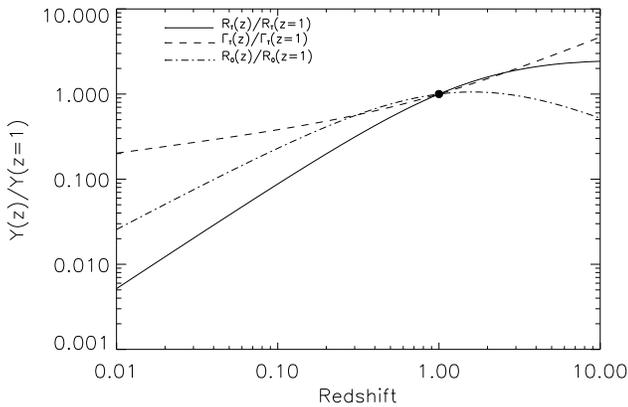,width=8.5cm}
\caption{Dependence on the assumed redshift of the estimated parameters with respect to the values computed assuming $z=1$. The solid line shows the transparency radius $R_{T}$, the dashed line the bulk Lorentz factor at transparency $\Gamma_{T}$ and the dot--dashed line the initial fireball radius $R_{0}$. }
\label{fg3}
\end{figure}
\begin{enumerate}
\item {\bf Redshift}. One possibility to reduce $R_0$ (i.e. making $R_{0}\sim 10r_{g}$) is to assume a different redshift. Fig.\ref{fg3}  shows how the three parameters $R_{0}$, $R_{T}$ and $\Gamma_{T}$ change assuming a different $z$. All the curves are normalized to the values of the parameters obtained assuming $z=1$.   $R_0$ (dot--dashed line in Fig.\ref{fg3}) could be reduced by a factor  of $\sim$1000 if the redshift  were $z<10^{-3}$.  

\item {\bf Efficiency}.  Another unknown parameter is the efficiency $\eta$. If the transparency is reached during the coasting phase, the estimate of $R_{0}$ depends on $\eta^{3/2}$ (equation \ref{eq5}); in the acceleration phase $\eta\sim1$ and constant. The values of $R_{0}$ shown in  Fig.\ref{fg2} were obtained assuming a typical value of $\eta=20\%$ (inferred from the modelling of GRB afterglows - e.g.  Panaitescu \& Kumar 2000). However, in class--III bursts (Axelsson et al. 2012; Guiriec et al. 2011, Guiriec et al. 2012) the BB component contributes at most 5\% of the total flux. Note that this could be also the situation of several \ba\ bursts (R04; Ryde 2005) which are fitted by a BB plus a power law.  
In class--I bursts the efficiency instead of 20\% could be correspondingly smaller $\eta\sim1$\%. With this value the estimate of $R_0$ decreases by a factor of $\approx$100. 

\end{enumerate}
How  such a low value of $\eta$ can be explained?
\begin{itemize}
\item {\it Hot fireball}. If most of the initial power is in the form of thermal photons (i.e. a non--magnetic fireball), the BB luminosity can be relatively low if (1) the transparency is reached during the coasting phase and if (2) no dissipation happens below the photosphere. Indeed, in this case most of the initial energy has been converted to kinetic energy so that the efficiency of the thermal emission can be $\eta_{BB}\sim$1\%. This scenario could justify the BB component in class--I, II and III bursts. However, for class--I events one should still explain the non detection of the non--thermal component,  which  can  be either absent or extremely dim with respect to the dominating thermal one. One possibility is that, despite the fact that most of the internal energy has been transformed into bulk motion, internal shocks are inefficient in class--I bursts. The internal shock efficiency for shells of equal mass is $\epsilon_{IS}=1-(\Gamma_f\Gamma_b)^{1/2}/(\Gamma_f+\Gamma_b)$ where $\Gamma_f$ and $\Gamma_b$ are the bulk Lorentz factors of the front (f) and back (b) shells. In order to produce a shock the shells should catch up during the coasting phase (i.e. $\Gamma_b>\Gamma_f$) and a shock efficiency of few percent can be achieved if $\Gamma_b/\Gamma_f\simeq$2. 
Fig.\ref{fg3} (bottom panel) shows that in class--I bursts, either $\Gamma$ is almost constant with time or  $\Gamma$ is a decreasing power law of time without a significant amplitude variability. Inefficient internal shocks could account for the absence of the non--thermal component in class--I GRBs. In these cases, however, most of the kinetic energy would be dissipated at the external shocks producing a bright afterglow emission. 

\item {\it Cold fireball.} 
If only a small fraction of the initial power is in thermal photons and the fireball is magnetic (as expected if  a large magnetic field is necessary to efficiently extract the BH rotational energy), the acceleration is governed by the magnetic field. In this scenario, either if the transparency is reached during the acceleration or the coasting phase, the thermal component luminosity can be small ($\eta_{BB}\sim$1\%). However, a bright non--thermal component can be produced through magnetic reconnection or internal shocks. While this scenario can account for class--III and II bursts (Hascoet et al. 2013), the non--thermal component should be  extremely dim (or even suppressed) in class--I bursts. If the fireballs  are still highly magnetic when they coast (Granot et al. 2012), the dynamic efficiency of internal shocks can be low (e.g. Mimica \& Aloy 2012) but still magnetic reconnection should produce a dominating non--thermal component which is absent in class--I bursts. 

\end{itemize}

We have discussed how  to reduce $R_0$ by changing independently the redshift $z$ or the efficiency $\eta$ but any combination of these two parameters could reduce $R_{0}$. These considerations apply to  the four GRBs with a pure blackbody spectrum presented in this work (i.e. 941023, 951228, 990413 and 100510) except GRB 930214. In this case, as shown in Fig.\ref{fg2}, the initial $R_0$ is only a factor a few larger than a fiducial $10r_g$. However, $R_0$ in GRB 930214 increases with time becoming 1000$r_g$ at the end of the burst unless the efficiency $\eta$ evolves with time in the opposite sense.

\subsection{``Re--born" fireball model}

An alternative interpretation of the large values of $R_0$ derived in \S6 could be that this radius does not represent the radius of 
the base of the outflow (which we have compared with 10$r_g$ in the previous section).

Within the standard model of GRBs, it might be possible that 
 at the radius of the star $R_{\star}\sim10^{10}$ cm, the fireball that has accelerated to mildly relativistic velocities, impacts on the material left over by the star. This material is at rest with respect to the expanding fireball. At such distances, the fireball is still highly opaque so that the encounter reconverts most of its kinetic energy into random internal energy, i.e. the fireball is re-created. The re--born fireball starts to reaccelerate at $R_{\star}$.  
Such a possibility has been studied first by Thompson (2006) and also applied by Ghisellini et al. (2007).

If the impact is efficient to nearly halt the fireball so that $\Gamma_{i}\sim1$ (where the subscript {\it i} stands for ``impact"),  the equations derived  for the standard case should still be valid. In this case, however, the radius $R_{0}$ should be compared with the  radius of the star $R_{\star}$ corresponding to the impact with the left over material. Indeed,  our estimates of $R_{0}$ are of the order of 10$^{9}$--10$^{10}$ cm (mid panel in Fig.\ref{fg2}). 

In this scenario, the fireball will reaccelerate up to the radius where it becomes transparent $R_{T}\sim 10^{12}$ cm (top panel in Fig.\ref{fg2}) and most likely it will be still accelerating at the transparency. As a consequence, most of the internal energy will be released as blackbody photons. The efficiency should be large in this case. This scenario could explain the absence of non--thermal emission  in class--I bursts and predicts a faint afterglow emission. 


\section{Conclusions}

In this paper we have presented the time resolved spectral analysis of the \fe\ GRB 100507 whose main peculiarity is the presence of a  blackbody (BB) spectrum throughout its duration. No sign of a non--thermal spectral component is found in this burst. GRB 100507 is the first detected by \fe\ with a BB spectrum throughout its duration and it is the fifth burst so far known with this characteristic (class--I events as defined in \S1). The other four events were all detected by \ba. Here, we have collected them and compared their properties. In GRB 100507, we find that the BB temperature $kT$  varies, between 30 and 40 keV, during the burst, at odd with the smooth  monotonic evolution of $kT$ in the other four class--I GRBs. Also the flux of the BB in GRB 100507 is not correlated with the temperature:  both the BB temperature and its flux  vary by a factor of 2 to 4 during the burst. 

Under the hypothesis that the BB radiation is produced by the fireball when it becomes transparent, we  derived for class--I GRBs the basic fireball parameters: the bulk Lorentz factor $\Gamma_T$ and radius $R_T$ at transparency and the  radius $R_0$ where the fireballs were created. The values of $R_T\sim10^{12}$ cm and $\Gamma_T\sim 80--200$ are reasonably in agreement with the standard fireball model, while the estimated $R_0$ appears much larger than the typical $r_g$ of a few solar mass black hole. 

This apparent inconsistency could be solved if redshift of these bursts is unlikely low  $z<0.01-0.001$ or, more likely the  fraction of thermal photons released at the photosphere is intrinsically small $\eta\sim1$\%. The latter could be the case of a hot fireball which becomes transparent in the coasting phase when most of the internal energy has been converted into kinetic energy or of a magnetically dominated fireball with an intrinsically small content of thermal photons. In both cases the non--thermal emission which could be produced by internal shocks or magnetic reconnection in the optically thin phase should be almost suppressed. Inefficient internal shocks in class--I bursts are supported by the almost constant or mildly decreasing (with small amplitude variations) $\Gamma_T$ values shown in Fig.\ref{fg2}.



An alternative interpretation of the large $R_0$ estimates is that this is the radius of the star surface where the fireball, launched by the central engine, encounters some left over material deposited before the jet launching by the progenitor star. Since the fireball at such distance $R_0\sim R_\star$ is still opaque, it reconverts most of its kinetic energy into internal energy and from that radius a new fireball is created which, starts to re--accelerate. 

\section*{Acknowledgements} 
We thank the referee for useful comments that improved the manuscript. 
We acknowledge the 2011 PRIN-INAF grant for financial support. 
AP thanks the Brera Observatory for the kind hospitality during his graduate work.

\end{document}